\documentstyle[12pt]{article}

\begin{document}

\title{\sc FAKE $R^{4}$'s, EINSTEIN SPACES
AND SEIBERG-WITTEN MONOPOLE EQUATIONS }
\author{Cihan Sa\c{c}l\i o\~{g}lu$^{1,2}$\\
\date{$^{1}$Physics Department, Bo\~{g}azi\c{c}i University \\
80815 Bebek--\.{I}stanbul, Turkey\\
and \\
$^{2}$Feza G\"{u}rsey Institute \\   Bo\~{g}azi\c{c}i
University--TUBITAK  \\ 81220 \c{C}engelk\"{o}y, {I}stanbul--
Turkey}} \maketitle
\begin{abstract}
We discuss the possible relevance of some recent mathematical
results and techniques on four-manifolds to physics.  We first
suggest that the existence of uncountably many $R^4$'s with
non-equivalent smooth structures, a mathematical phenomenon unique
to four dimensions, may be responsible for the observed
four-dimensionality of spacetime. We then point out the remarkable
fact that self-dual gauge fields and Weyl spinors can live on a
manifold of Euclidean signature without affecting the metric.  As
a specific example, we consider solutions of the Seiberg-Witten
Monopole Equations in which the $U(1)$ fields are covariantly
constant, the monopole Weyl spinor has only a single constant
component, and the 4-manifold ${\cal M}_{4}$ is a product of two
Riemann surfaces $\Sigma_{p_{1}}$ and $\Sigma_{p_{2}}$. There are
$p_{1} -1(p_{2} - 1)$ magnetic(electric) vortices on
$\Sigma_{p_{1}}(\Sigma_{p_{2}})$, with $p_{1} + p_{2} \geq 2$
($p_{1} = p_{2}= 1$ being excluded). When the two genuses are
equal, the electromagnetic fields are self-dual and one obtains
the Einstein space $\Sigma_{p}\times\Sigma_{p}$~, the monopole
condensate serving as the cosmological constant.
\end{abstract}

\pagebreak

\noindent  {\bf 1.  Introduction:}

Some time after the discovery of uncountably many four-manifolds
that are homeomorphic but not diffeomorphic to $R^4$
\cite{Gompf},\cite{Taubes}, Atiyah \cite{Atiyah} challenged
physicists to find a physical realization for these 'Fake
$R^{4}$'s'. Our first aim is to suggest a possible response to
this challenge. We will argue in the next section that spacetime
may be four-dimensional because of this uncountable profusion of
Fake $R^{4}$'s, as opposed to only a single standard $R^n$ in any
other dimension $n$.

In the rest of the note we examine the relationship between the
Seiberg-Witten monopole equations (SWME) \cite{Witten} and
physics.  The physics interest in solutions of Euclidean
matter-coupled gravity has to do with semiclassical treatments of
vacuum tunnelling, estimating the path integral of gravity and
possibly for understanding the earliest phases of the universe.
Some mathematical questions one may raise about such manifolds
are:  (i) which smooth compact manifolds admit Einstein metrics,
and, (ii) how many (modulo diffeomorphisms and rescalings)
different Einstein metrics can be put on the same manifold
\cite{LeBrun}.  In ref. \cite{LeBrun} global methods are employed.
Our approach, in contrast, is based on explicit local equations
and their solutions; these provide a pedestrian route to Einstein
spaces and $Spin_c$ structures.  Since most of these results have
already been presented elsewhere \cite{Sacli}, the discussion will
be rather condensed.

~~~~~~~~~~~~~~~~~~~~~~~~~~~~~~~~~~~~~~~~~~~~~~~~~~~~~~~~~~~~~~~~~~~~~~~

~~~~~~~~~~~~~~~~~~~~~~~~~~~~~~~~~~~~~~~~~~~~~~~~~~~~~~~~~~~~~~~~~~~~~~~~

\noindent {\bf 2. Fake $R^{4}$'s and spacetime:}

The idea that the dimensionality of spacetime can be derived from
some dynamical principle rather than being accepted as an a priori
fact is a feature of String theory, so it may be instructive to
examine how this is handled in the Polyakov path integral
formulation of strings. The measure consists of
$[dX^\mu][g_{ab}]$, but it is customary to treat target space
coordinates and the world sheet very differently. For the latter,
care is taken to include (i) all possible world sheet topologies
(by performing a sum over genera), (ii) all inequivalent complex
structures (by integrating over moduli space).  On the other hand,
for no obvious or compelling reason, similar considerations are
not extended to target space, a.k.a. the spacetime manifold.  A
critical upper limit for the dimension of spacetime emerges
indirectly when one demands cancellation between the ghost and
target space contributions to the anomaly, but this is clearly
different from the deliberate choices one allows for the world
sheet. One then appeals to the possibility of space being curved
in General Relativity and argues that some as yet not understood
physical mechanism closes all but three space dimensions onto
themselves.

Here we would like to propose tentatively another possible
physical mechanism leading to a four-dimensional spacetime.
Basically, the idea is to let a generalized path integral decide
the dimensionality of the target spacetime. A path integral for
quantizing Einstein's theory involving a sum over various
topologies has already been considered \cite{HawkingPerry} .  In
similar fashion, imagine extending the String theory path integral
by allowing a sum over all possible target space manifolds.  In
partial analogy with what is done with the world sheet, the sum
could be performed as follows: Choose a particular dimension $D$,
sum over all possible topologies at that dimension just as one
sums over the sphere, the torus, etc. We note at this point that
the physical signatures of different global 'Cosmic Topologies'
for four-dimensional spacetime have already been examined
\cite{Frenchies}.  Next, when the dimension admits them, sum over
inequivalent differentiable structures. When finished with that
particular dimension, move on to $D+1$. Admittedly, the sum cannot
be made precise without knowledge, among other things, of the
Boltzmann factors to be assigned to the different differentiable
structures. However, a very gross fact at $D=4$ renders this
knowledge superfluous if the Boltzmann factors of these structures
are comparable.  The uncountable number of $R^4$'s, all but one of
which are "fake", completely overwhelms all other contributions
from other dimensions and topologies. Since fake and standard
$R^4$'s cannot be distinguished by looking at a local
neighborhood, but only by comparing the overall global structure
 \cite{LANGL}, the argument would suggest that the familiar local $R^4$ we
live in is in fact some average of the standard one and all its
fakes.  The differentiable structure being independent of the
metric, the above considerations hold regardless of the signature
of the $R^4$ metric.

~~~~~~~~~~~~~~~~~~~~~~~~~~~~~~~~~~~~~~~~~~~~~~~~~~~~~~~~~~~~~~~~

~~~~~~~~~~~~~~~~~~~~~~~~~~~~~~~~~~~~~~~~~~~~~~~~~~~~~~~~~~~~~~~~

\noindent  {\bf 3. Einstein spaces and Seiberg-Witten monopole
equations:}

We start by some very general facts, which, however, lead to
surprising conclusions about physics on Euclidean manifolds.
Recall that as we go from $(-+++)$ signature to $(++++)$, the
Dirac bilinear covariants $\psi^{\dag}\gamma^0\Gamma \psi$ turn
into $\psi^{\dag}\Gamma \psi$. In a Weyl representation, the
$\gamma^{\mu}$ are block-off diagonal, while $\gamma^5$ is block
diagonal; they are all Hermitean.  We may take, for example,
$\gamma^i=\tau_1\otimes\sigma_i$, $\gamma^4=\tau_2\otimes I$ and
$\gamma^5=\tau_3\otimes I$. This means that in Euclidean signature
the vector and the axial vector currents and the energy-momentum
tensor for a Weyl spinor automatically vanish! Now these are
precisely the bilinears which would have served as sources for the
gauge field strength tensor, its dual, and the Einstein tensor,
respectively. Hence the gauge fields and the metric are unaffected
by the Weyl fields, and the field strengths are covariantly
constant.

Euclidean signature also allows real, non-trivial self-dual or
anti-self-dual gauge fields. These automatically solve the
sourceless Maxwell's equations. Furthermore, the energy-momentum
tensor for these fields also vanishes, and we have the
\textit{vacuum} Einstein field equations

\begin{equation}\label{1}
{\cal R}_{\mu\nu}-\frac{1}{2}g_{\mu\nu}{\cal R}= \Lambda
g_{\mu\nu},
\end{equation}

\noindent where we have allowed a cosmological constant $\Lambda$.
This is just the defining equation for the Einstein spaces
considered in \cite{LeBrun}.  What we have said so far holds for
all gauge fields, whether Abelian or not, and also for Weyl
spinors which are not necessarily singlets under the gauge group.
However, from now on, we will work with a $U(1)$ gauge field
$A_{\mu}$ and a single Weyl spinor $\psi$, which then satisfy the
Dirac equation

\begin{equation}\label{2}
\not{\! \! D}_{A} \psi = \gamma^{a} E^{\mu}_{a}
(\partial_{\mu}+iA_{\mu}+\frac{1}{8}
\omega^{bc}_{\mu}[\gamma_{b},\gamma_{c}])\psi=0,
\end{equation}

\noindent where $\omega^{bc}_{\mu}$ are the spin connection
coefficients and $E^{\mu}_{a}$ the inverses of the vierbeins.

If one assumes that the Weyl spinor represents a massless
monopole, the above are exactly the fields in the SWME. The SWME
have proven to be a considerably more practical tool for
distinguishing inequivalent smooth structures on homeomorphic
four-manifolds than the earlier Donaldson theory (the recent
realization that \cite{Park}, \cite {Wang}
Seiberg-Witten-Donaldson theory still fails to detect certain
inequivalent smooth structures does not affect the present
analysis). The SWME consist of (\ref{2}) and

\begin{equation}\label{3}
F^{+}_{\mu\nu}\equiv\frac{1}{2}(F_{\mu\nu} +
\frac{1}{2}\epsilon_{\mu \nu \alpha \beta}F^{\mu\nu})
=-\frac{i}{4} \psi^{\dag}[\gamma_{\mu},\gamma_{\nu}]\psi~~~.
\end{equation}

\noindent The SWME have at least one solution $(A_{\mu}, \psi)$
for every metric $g_{\mu\nu}$ on the manifold \cite{LeBrun}. This
is related to the mathematical fact that even when a 4-manifold
does not admit a spin structure because of an obstruction in the
form of a spinor phase ambiguity, the introduction of a $U(1)$
bundle can compensate for the phase mismatch. One then has a
$Spin_c$ structure \cite{Egu} and spinors are allowed
\cite{Morgan}. Using the Weitzenbock-Lichnerowicz formula, one can
further show that (\ref{2}) and (\ref{3}) do not admit non-trivial
non-singular solutions unless the scalar curvature ${\cal R}$ is
negative somewhere \cite{Akb}. If one now tries to find
simultaneous solutions of (\ref{1}), (\ref{2}), (\ref{3}) and
$F_{\mu\nu}=\widetilde{F_{\mu \nu}}$, it is clear that only
constant negative curvature Einstein spaces will admit
non-singular matter fields. Putting these facts together, we see
that all possible simultaneous non-singular, non-trivial solutions
of the SWME and (\ref{1}) require self-dual gauge fields and
constant negative curvature. We will give explicit examples of
these in the later sections. Semi-trivial solutions without
spinors and anti-self-dual fields are also a possibility, as are
non-trivial but singular solutions with self-dual gauge fields.

\noindent  {\bf 4. Explicit solutions of the form ${\cal
M}^{(1)}_{2}\times{\cal M}^{(2)}_{2}$:}

As an example which permits simple explicit calculations, we look
for solutions of the SWME in the form of a product of two
two-manifolds parametrized by coordinates $(x^{1}, x^{2})$ and
$(x^{3}, x^{4})$; it is then natural to assume that in $(A^{1},
A^{2}, A^{3}, A^{4})$ the first two components depend on the first
pair of coordinates and the last two on the second pair.  The SWME
then imply that $\psi$ has only one non-vanishing component, say
$\psi_1$, which is forced to be a constant.  The scalar curvature
$R=2R^{12}_{12}+2R^{34}_{34}$ turns out to have the value $ -
2|\psi_{1}|^{2}$, so the solutions are guaranteed to be
non-singular. Introducing another constant $|\phi|$, we can break
up $R$ into the curvatures of ${\cal M}^{(1)}_{2}$ and ${\cal
M}^{(2)}_{2}$ by writing $R^{12}_{12}= -|\phi|^2$ and
$R^{34}_{34}= -(|\psi_{1}|^{2}-|\phi|^{2})$.  We then see that
there are three kinds of solutions in which (i) both manifolds
have constant negative curvature $(|\psi_{1}|>|\phi|)$, (ii) one
is flat and the other has constant negative curvature
$(|\psi_{1}|=|\phi|)~ or~ |\phi|=0)$, and (iii) one has positive
and the other constant negative curvature $(|\psi_{1}|<|\phi|)$,
but the total scalar curvature is still negative. In terms of
dimensionless complex coordinates $z^1\equiv \sqrt{2}|\phi|(x^1 +
i x^2)$ and $z^2\equiv \sqrt{2}(|\psi_{1}|^2 -
|\phi|^2)^{\frac{1}{2}}(x^3 + i x^4)$, the solution for case (i)
is

\begin{equation}\label{4}
\omega^{1}_{2}= - i \{ \frac{1}{2}d\ln (\frac{d\overline{g}_{1}}
{d\overline{z}_{1}} \frac {dz_{1}}{dg_{1}}) + \frac
{(g_{1}d\overline{g}_{1}- \overline{g}_{1}dg_{1})} {(1-
g_{1}\overline{g}_{1})} \},
\end{equation}
\begin{equation}\label{5}
\omega^{3}_{4} = - i \{ \frac{1}{2}d\ln (\frac{d\overline{g}_{2}}
{d\overline{z}_{2}} \frac {dz_{2}}{dg_{2}}) + \frac
{(g_{2}d\overline{g}_{2}- \overline{g}_{2}dg_{2})} {(1-
g_{2}\overline{g}_{2})} \},
\end{equation}
\begin{equation}\label{6}
A = - \frac{1}{2}(\omega^{1}_{2}+\omega^{3}_{4}),
\end{equation}
\begin{equation}\label{7}
R^{1}_{2} = - 2i \frac {dg_{1} \wedge d\overline{g}_{1}} {(1-
g_{1}\overline{g}_{1})^{2}},~~~R^{3}_{4} = - 2i \frac {dg_{2}
\wedge d\overline{g}_{2}} {(1- g_{2}\overline{g}_{2})^{2}},
\end{equation}
\begin{equation}\label{8}
F= i \frac {dg_{1} \wedge d\overline{g}_{1}} {(1-
g_{1}\overline{g}_{1})^{2}}+ i \frac {dg_{2} \wedge
d\overline{g}_{2}} {(1- g_{2}\overline{g}_{2})^{2}}= -
\frac{1}{2}(R^{1}_{2} +R^{3}_{4}),
\end{equation}
\begin{equation}\label{9}
ds^{2}({\cal M}^{(1)}_{2})=
\frac{dg_{1}d\overline{g_{1}}}{(1-g_{1}\overline{g_{1}})^{2}},
\end{equation}
\begin{equation}\label{10}
ds^{2}({\cal M}^{(2)}_{2})=
\frac{dg_{2}d\overline{g_{2}}}{(1-g_{2}\overline{g_{2}})^{2}},
\end{equation}

\noindent where $g_1 (z_1)$ and $g_2 (z_2)$ are arbitrary analytic
functions originating from Liouville equations, which in turn were
obtained from the SWME using our Ansatz.

We will illustrate case (ii) by taking $|\phi|=0$, which
corresponds to a flat metric for the first manifold, and one of
constant negative curvature for the second.  The solutions for the
first manifold are then no longer characterized by a solution of
the Liouville equation, but instead by harmonic functions, which
we will later take as the real part of an analytic function
$h(z_1)$. The results then follow from (\ref{4})-(\ref{10}) by
setting $g_1=0$ except in (\ref{9}), where the metric is now flat;
note also the first two components of the $U(1)$ connection are
absent.

In the final case (iii), we take the second manifold as the one
with positive curvature and the solution is given by

\begin{equation}\label{11}
\omega^{1}_{2}= - i \{ \frac{1}{2}d\ln (\frac{d\overline{g}_{1}}
{d\overline{z}_{1}} \frac {dz_{1}}{dg_{1}}) + \frac
{(g_{1}d\overline{g}_{1} - \overline{g}_{1}dg_{1})} {(1-
g_{1}\overline{g}_{1})} \},
\end{equation}
\begin{equation}\label{12}
\omega^{3}_{4} = i \{ \frac{1}{2}d\ln (\frac{d\overline{g}_{2}}
{d\overline{z}_{2}} \frac {dz_{2}}{dg_{2}}) + \frac
{(g_{2}d\overline{g}_{2}- \overline{g}_{2}dg_{2})} {(1+
g_{2}\overline{g}_{2})} \},
\end{equation}
\begin{equation}\label{13}
A = - \frac{1}{2}(\omega^{1}_{2}+\omega^{3}_{4}),
\end{equation}
\begin{equation}\label{14}
R^{1}_{2} = - 2i \frac {dg_{1} \wedge d\overline{g}_{1}} {(1-
g_{1}\overline{g}_{1})^{2}},~~~R^{3}_{4} = +2i \frac {dg_{2}
\wedge d\overline{g}_{2}} {(1+ g_{2}\overline{g}_{2})^{2}},
\end{equation}
\begin{equation}\label{15}
F= i \frac {dg_{1} \wedge d\overline{g}_{1}} {(1-
g_{1}\overline{g}_{1})^{2}} - i \frac {dg_{2} \wedge
d\overline{g}_{2}} {(1+ g_{2}\overline{g}_{2})^{2}}= -
\frac{1}{2}(R^{1}_{2} +R^{3}_{4})~,
\end{equation}
\begin{equation}\label{16}
ds^{2}({\cal M}^{(2)}_{2})=
\frac{dg_{2}d\overline{g_{2}}}{(1+g_{2}\overline{g_{2}})^{2}}.
\end{equation}

The equations (\ref{4})-(\ref{16}) show that points on the
manifold are projected stereographically onto the $g_1$ and $g_2$
complex planes. The topological properties of the manifolds are
encoded in the functions $g_1, g_2$ and $h$. The simplest cases
are $g_1=z_1, g_2=z_2$, giving a product of two hyperboloids in
case (i), and a hyperboloid and a sphere in case (iii). Similarly,
in case (ii), the simplest version of the flat metric
$ds^2=|\partial h(z_1)|^2 dz_1 \overline{dz_1}$ is for
$h(z_1)=z_1$; the second manifold is again a hyperboloid. However,
if we choose $h$ to be an inverse elliptic function, ${\cal
M}^{(1)}_{2}$ is turned into $T^2$. Similarly, the hyperboloid
with coordinates $g_1(z_1)$ becomes the Riemann surface
$\Sigma_{p_{1}}$ of genus $p_1$ when $g_1(z_1)$ is the Fuchsian
function \cite{Nehari} (used in uniformizing an algebraic function
defined on the same Riemann surface) because the Fuchsian function
tessellates the hyperboloid into $4p_{1}$-gons with geodesic edges
identified in the standard way \cite{Dubrovin}. Finally, in case
(iii), ${\cal M}^{(2)}_{2} =S^2$, and no such topologically
different variants are possible. In this case, the most general
one-to-one mapping of the Riemann sphere to itself is of the form

\begin{equation}\label{17}
g_{2}(z_{2}) =  \frac{az_{2} + b}{cz_{2} + d}.
\end{equation}

Summarizing, we have found explicit solutions of the SWME in the
form $\Sigma_{p_{1}} \times \Sigma_{p_{2}}$, with the constraint
that $p_{1} + p_{2} \geq 2$, excluding $p_1 = p_2 = 1$.
Physically, $p_1 - 1 (p_2 - 1)$ is the number of
magnetic(electric) vortices on ${\cal M}^{(1)}_{2} ({\cal
M}^{(2)}_{2})$.

Are these solutions unique? This can be partially answered by
computing the virtual dimension $W$ of the moduli space of
solutions defined by $- \frac{(2\chi + 3 \sigma)}{4} + c_{1}^{2}$.
For our solutions, it is easy to show that the signature $\sigma =
0$, the Euler characteristic $\chi = 4(p-1 - 1 )(p_2 - 1) =
2c_1^2$~; hence $W=0$.  The solutions thus correspond to a
discrete set of points in moduli space.

\noindent  {\bf 5. Einstein spaces of the form $\Sigma_{p} \times
\Sigma_{p}~, p\geq 2:$}

Of the solutions so far displayed, the ones with covariantly
constant self-dual $U(1)$ fields are guaranteed also to solve the
coupled Einstein-Maxwell-Dirac equations by the discussion in
section 3.  A glance at these solutions shows that the $U(1)$
connection over each manifold is just minus one-half of the spin
connection for the same manifold; self duality thus makes the two
manifolds identical, with $p-1$ magnetic vortices on the first and
the same number of electric vortices on the other.  The total
scalar curvature $R= - 2 |\psi_1|^2$ by the SWME and this shows
that the cosmological constant $\Lambda$ consists of the constant
massless magnetic monopole condensate.  These solutions may be
regarded as an extension of the Bertotti-Robinson
\cite{Ber},\cite{Rob} solutions, with Weyl spinors added, to
Euclidean signature. Explicit computation, both analytical, and by
using REDUCE, shows that none of the other SWME solves the
Einstein-Maxwell-Dirac equations.

What would have happened had we chosen $|\psi_1| = 0, |\psi_2|\neq
0$ in the beginning? Examining the formulae, it is easy to see
that this replaces complex variables with their complex
conjugates, changes the sign of the $U(1)$ fields but leaves the
metric unchanged. Recalling that the two components of a Weyl
spinor correspond to particles and antiparticles of the
\textit{same} handedness, we see that the charge conjugate of the
original solution is obtained.

\noindent {\bf Acknowledgements}

I am grateful to S. Akbulut for private instruction in
Seiberg-Witten theory, to P. Argyres for clarifying a point
involving the virtual dimension, and to Y. Nutku for informing me
of \cite{Ber} and \cite{Rob}, as well as for the REDUCE check. I
thank R. P. Langlands for observing that fake and standard $R^4$'s
are locally indistinguishable, M. Ar\i k, T. Dereli and R.
G\"{u}ven for useful discussions, and U. Kayserilioglu for help
with LATEX.

\end{document}